\documentclass[12pt]{article}

\newif\iffigs\figstrue

\usepackage{epsfig,latexsym}
\usepackage{amsmath}
\usepackage{color}
\usepackage{graphicx}
\usepackage{verbatim}
\usepackage{mathrsfs}
\usepackage{amssymb}

\DeclareMathAlphabet{\mathpzc}{OT1}{pzc}{m}{it}

 \csname
@addtoreset\endcsname{equation}{section}

\def\gz0{\gamma^{0}}

\def\bec{\begin{center}}
\def\ec{\end{center}}

\def\12{\frac{1}{2}}

\def\DH{\rm I\kern-1.5pt\rm H\kern-1.5pt\rm I}

\def\DR{\rm I\kern-1.45pt\rm R}
\def\DC{\kern2pt {\hbox{\sqi I}}\kern-4.2pt\rm C}

\newcommand{\bF}{{\overline F}{}}

\newcommand{\bP}{{\overline P}}
\newcommand{\bV}{{\overline V}{}}

\newcommand{\I}{{\rm i}}

\newcommand{\beq}{\begin{equation}}
\newcommand{\eeq}{\end{equation}}
\newcommand{\bea}{\begin{eqnarray}}
\newcommand{\eea}{\end{eqnarray}}
\newcommand{\bi}{\begin{itemize}}
\newcommand{\ei}{\end{itemize}}

\newcommand{\p}[1]{(\ref{#1})}

\renewcommand{\Im}{{\tt Im\,}}
\renewcommand{\Re}{{\tt Re\,}}

\usepackage{slashed}

\usepackage{float}

\usepackage{caption}

\usepackage{epsf}
\usepackage{amsmath}

\usepackage{multirow}

\usepackage{epsfig}
\usepackage{cite}
\usepackage{color,colordvi}

\newcounter{hran}

\makeatletter
\renewcommand\section{\@startsection {section}{1}{\z@}%
                               {-3.5ex \@plus -1ex \@minus -.2ex}%
                               {2.3ex \@plus.2ex}%
                               {\normalfont\large\bfseries}}
\makeatother

\vspace{0.5cm}

\begin{document}
\thispagestyle{empty}

\begin{flushright}
CERN-PH-TH/2016-017\\
\end{flushright}

\vspace{15pt}

\begin{center}


{\Large\sc Two--Field Born--Infeld with Diverse Dualities}\\


\vspace{35pt}
{\sc S.~Ferrara${}^{\; a,b,c}$, A.~Sagnotti${}^{\; d}$ and A.~Yeranyan${}^{\; e,b}$}\\[15pt]

{${}^a$\sl\small Department of Theoretical Physics\\
CH - 1211 Geneva 23, SWITZERLAND \\ }
e-mail: {\small \it sergio.ferrara@cern.ch}
\vspace{6pt}

{${}^b$\sl\small INFN - Laboratori Nazionali di Frascati \\
Via Enrico Fermi 40, I-00044 Frascati, ITALY}\vspace{6pt}

{${}^c$\sl\small Department of Physics and Astronomy \\ U.C.L.A., Los Angeles CA 90095-1547, USA}\vspace{6pt}

{${}^d$\sl\small
Scuola Normale Superiore and INFN\\
Piazza dei Cavalieri \ 7\\I-56126 Pisa \ ITALY \\}
e-mail: {\small \it sagnotti@sns.it}\vspace{6pt}

{${}^e$\sl\small Centro Studi e Ricerche Enrico Fermi\\
Via Panisperna 89A, 00184, Roma, Italy\\}
e-mail: {\small \it ayeran@lnf.infn.it}

\vspace{8pt}

\vspace{25pt} {\sc\large Abstract}
\end{center}

\noindent We elaborate on how to build, in a systematic fashion, two--field Abelian extensions of the Born--Infeld Lagrangian. These models realize the non--trivial duality groups that are allowed in this case, namely $U(2)$, $SU(2)$ and $U(1) \times U(1)$. For each class, we also construct an explicit example. They all involve an overall square root and reduce to the Born--Infeld model if the two fields are identified, but differ in quartic and higher interactions. The $U(1) \times U(1)$ and $SU(2)$ examples recover some recent results obtained with different techniques, and we show that the $U(1) \times U(1)$ model admits an ${\cal N}=1$ supersymmetric completion. The $U(2)$ example includes some unusual terms that are not analytic at the origin of field space.
\vskip 36pt
\begin{center}
{\sl Contribution to a special Nucl.~Phys.~B issue dedicated to the memory of Raymond Stora}
\end{center}
\vfill

\noindent

\baselineskip=20pt

\pagebreak

\newpage

\setcounter{page}{2}
\setcounter{equation}{0}
\section{Introduction} \label{sec:intro}
The Born--Infeld (BI) theory \cite{BI} is described by the Lagrangian
\beq\label{BI}
{\cal L} \ =\ f^2 \left[\ 1 \ -\ \sqrt{1\ +\ \frac{1}{2\,f^2}\,\left({\cal F}\cdot {\cal F}\right) \ -\ \frac{1}{16\,f^4}\,\left({\cal F}\cdot\widetilde{ {\cal F}}\right)^2}\ \right] \ ,
\eeq
where ${\cal F}_{mn}$ is an electromagnetic field strength in the standard four--component notation. It
was initially put forward as an elegant refinement, based on the determinant of $\eta_{mn} + \frac{1}{f}\, {\cal F}_{mn}$, of an earlier proposal \cite{Born} enforcing a dynamical upper bound on the electric field of a point charge. Both Lagrangians involve a square root, and both models do entail the same dynamical bound, much as occurs for the speed in Special Relativity. However, the choice of eq.~\eqref{BI} is particularly interesting, precisely due to the last term inside the square root. Schr\"odinger soon noticed \cite{SCHRO}, indeed, that the non--linear BI field equations afford a subtle and surprising realization of electric--magnetic duality in an interacting system, or if you will in a non--linear relativistic medium.

The BI theory made a striking and unexpected comeback in String Theory \cite{stringtheory}, in the 1980's, when Fradkin and Tseytlin \cite{FT} first linked it to the dynamics of open strings in a constant electromagnetic background. The phenomenon is of utmost interest, since it is an exact manifestation of the deformed spectra \cite{ACNY} of $D$--branes \cite{polchinski}, the extended objects that populate orientifold vacua \cite{orientifolds}.
However, two types of corrections affect it. The first is the generic presence of interactions involving derivatives of ${\cal F}_{mn}$, while the second is the non--abelian extensions that manifest themselves when $D$-branes are superposed. Both types of effects are unfortunately not fully understood (for a review see \cite{tseytlin_rev}), but the BI theory remains an important benchmark for all these searches.

A second, related reason of interest on the BI theory, has to do with the partial breaking of supersymmetry. When completed by the addition of gaugino interactions \cite{deser,CF}, the model of eq.~\eqref{BI} conceals indeed a second, non--linearly realized supersymmetry \cite{BG,RT,FPSSY,FPSSY2,DAURIA,BELLUCCI}, while $f$ defines the supersymmetry breaking scale. The superspace formulation rests on ${\cal N}=2$ constrained superfields, much along the lines of what happens for the Volkov--Akulov model \cite{VA}, and thus for the ${\cal N}=1 \to {\cal N}=0$ breaking \cite{constrained}. This is nicely consistent with the link between BI and D-branes, where partial breaking found originally a proper setting \cite{POL_PARTIAL}. Partial breaking of supersymmetry affords an alternative realization in models of ${\cal N}=1$ global supersymmetry with non--renormalizable \cite{APT} (magnetic) superpotential terms and Fayet--Iliopoulos \cite{FI} terms. In decoupling limits, one can recover multi--field extensions of the BI theory depending of $N$ field strengths ${{\cal F}^i}_{mn}$, $i=1,..N$ with \cite{FPSSY,FPSSY2,DAURIA,BELLUCCI}: they involve in general multiple square--roots and have been fully classified up to cases with three gauge fields. It would be interesting to clarify their possible links with D-branes. At any rate, in these multi-field models the duality group does not extend beyond the BI case.

Eventually one would like to extend the BI construction to Supergravity \cite{supergravity}, which would provide a low--energy characterization of D-brane systems with (partially) broken supersymmetry. D-branes typically do bring along, in general, non--linear realizations of supersymmetry \cite{dm}, since for one matter their presence in the vacuum breaks some translational symmetries, which affect their low--energy modes via shifts of scalars. The coupling of constrained multiplets to Supergravity \cite{constrained_sugra} has led to a resurgence of these ideas \cite{bsb_new}, also in connection with ``brane supersymmetry breaking'' \cite{bsb} and the KKLT construction \cite{KKLT}. In this case supersymmetry is fully broken, but it is extremely important to explore and characterize similar types of systems allowing for partial breaking, in various dimensions. All in all, the non--BPS combinations of BPS objects of brane supersymmetry breaking are possibly the simplest entry point into the intricate dynamics of non--supersymmetric brane systems (for a recent review see \cite{POL15}).

As we have anticipated, a key missing ingredient of present constructions are the generalized electric--magnetic dualities that play a central role in extended supergravity \cite{FSZ}. Duality symmetries for systems of Abelian field strengths were characterized, in general, by Gaillard and Zumino (GZ) \cite{GZ}. Drawing some inspiration from \cite{FSZ,CJ}, they showed that, with $N$ Abelian field strengths ${{\cal F}^i}_{mn}$ the maximal possible duality group is $U(N)$, which can extend at most to $Sp(2N,R)$ in the presence of scalars, and is accompanied by chiral rotations if fermions are present. The simplest example of this type is pure ${\cal N}=4$ supergravity, whose duality group is $SU(4) \times Sp(2,R)$, where the latter also acts on the axion--dilaton system \cite{CSF}. These results were analyzed in depth and extended in a number of works following \cite{GZ}, which include \cite{subsequent,KT}. The GZ formulation also raised the natural question of building corresponding extensions of the BI theory. The problem was set up in general in \cite{ABMZ}, but no analytic solutions were found for $N>1$. Non linear deformations of ${\cal N}=2$ Electrodynamics that are $U(1)$ duality invariant were also investigated. However, they  were not proven to be non--linear realisations of a higher ${\cal N}=4$ supersymmetry Ref. \cite{KT,extra}.

During the last decade, Ivanov and Zupnik (IZ) were responsible for a major independent line of development, which rests on the combined use of master actions and tensor auxiliary fields \cite{BIK1,Iv}. Master actions combining field strengths and their duals are a familiar tool to investigate electric--magnetic dualities, and in connection with scalar auxiliary fields they make Legendre transforms simple and elegant for the BI theory \cite{RT}. While duality transformations mix, in general, the field strengths ${{\cal F}^i}_{mn}$ and their duals ${{\cal G}^i}_{mn}$, which are non--linear functions of them, the IZ tensor auxiliary fields transform \emph{linearly} under dualities, \emph{in a universal way that is independent of the dynamics}. All \emph{bona fide} interactions that are duality invariant can be expressed solely in terms of them, which makes a systematic search for extended dualities possible. However, the reversal to the ordinary field strengths is typically difficult, and thus no simple closed--form multi--field examples were found.

In this paper we build, along the lines traced by IZ, three prototype analytic extensions of the BI model involving two field strengths ${{\cal F}^i}_{mn}$ $(i=1,2)$ that realize the possible extended duality groups, namely $U(2)$, $SU(2)$, $U(1) \times U(1)$. All these models reduce to the BI theory when the two field strengths are identified. The $U(2)$ model is new, but includes a peculiar term that is not analytic at the origin of field space, while the others reproduce results that we had previously presented in \cite{FSY}. In the weak--field limit, all these models
reduce to two copies of the Maxwell theory. Moreover, they all rest on one and the same expression in terms of auxiliary variables, which emerges naturally and is essentially the same that, for $N=1$, determines the BI theory. For more than two fields we have not found, so far, examples of comparable simplicity.

The plan of the paper is as follows. In Section \ref{sec:2} we review the previous construction \cite{BIK1,Iv} of models with a single field strength. In Section \ref{sec:2.2} we present a one--parameter deformation of the BI theory that is also invariant under $U(1)$ duality and contains some contributions that are not analytic at the origin of field space. In Section \ref{sec:3} we turn to the two--field case, and the following subsections describe the construction of our three prototype examples, with duality groups $U(2)$, $SU(2)$ and $U(1) \times U(1)$. The first model contains non--analytic terms that are akin to those met in Section \ref{sec:2.2}. In Section \ref{sec:susy} we discuss the possibility of extending the prototype models in order to accommodate ${\cal N}=1$ supersymmetry. Finally, Section \ref{sec:4} contains some concluding remarks.

\section{One--Field Models: BI Theory and a Family of Extensions} \label{sec:2}

Master actions combining field strengths with their duals are a familiar tool to approach dualities via Legendre transforms, but they can be very useful also to address the solution of the GZ constraints \cite{GZ} and the continuous duality symmetries of field equations.

The approach that will concern us here originates from the work of IZ \cite{BIK1,Iv}. Their key step was the introduction of tensorial counterparts $V_{\alpha\beta}$ and $\overline{V}_{\dot\alpha\dot\beta}$ of the Maxwell field strengths $F_{\alpha\beta}$ and $\bF_{\dot\alpha\dot\beta}$. We shall adopt this two--component notation to a large extent, reserving to Section \ref{sec:4} the translation of final results into the four--component form.

The authors of \cite{BIK1} first considered the redefinitions
\beq\label{eq1}
F_{\alpha\beta}\ = \ \left(\frac{1+\bV^2}{1-V^2\, \bV^2}\right) \, V_{\alpha\beta}\,, \quad
\bF_{\dot\alpha \dot\beta}\ = \ \left( \frac{1+V^2}{1-V^2 \,\bV^2}\right)\, \bV_{\dot\alpha \dot\beta}\ .
\eeq
Also in view of the following sections, let us define the scalar quantities
\bea
&&\phi\ = \ F^2\ , \quad \overline \phi\ = \ \overline F^2\ , \label{phi_phibar} \\
&&\nu\ =\ V^2\ ,\quad \overline \nu\ =\ \overline V^2\ , \quad a\ =\ \overline \nu\, \nu\ . \label{nu_nubar_a}
\eea
The first two involve $F_{\alpha\beta}$ and its complex conjugate, while the others involve the auxiliary field $V$. Lorentz invariance constrains the Lagrangian to depend on the variables of eq.~\eqref{phi_phibar}, and the standard BI action reads
\beq\label{BI11}
{\cal S}_{BI}\ =\ \int d^4x \left[ \,1\ -\ \, \sqrt{\frac{1}{4}\,\left( \phi\ -\ \overline{\phi}\right)^2\ +\ \left(\phi\ +\ \overline{\phi}\right)\ +\ 1}\, \right] \ .
\eeq
Interestingly, however, the redefinitions of eq.~\eqref{eq1} result in the far simpler, rational form
\beq\label{BI12}
{\cal S}_{BI}\ = \ -\ 2\ \int d^{\,4}x \, \frac{\Re[\nu] \ + \ a}{1 \ - \ a}\ ,
\eeq
an expression that will recur in the following sections.

Schr\"odinger readily noticed \cite{SCHRO} that the BI field equations
\beq
\partial_{\beta\dot\alpha}\, P_\alpha^\beta \ - \ \partial_{\alpha\dot\beta}\, \overline{P}_{\dot\alpha}^{\dot\beta} \ = \ 0\   \ , \qquad \label{EoM1}
\eeq
where
\beq
P^{\,\alpha\beta}(x) \ = \ i \ \frac{\delta {\cal S}}{\delta F_{\alpha\beta}(x)} \ , \qquad \overline{P}^{\,\dot{\alpha}\dot{\beta}}(x) \ = \ - \ i \ \frac{\delta {\cal S}}{\delta \overline{F}_{\dot{\alpha}\dot{\beta}}(x)}  \label{PfromS}
\eeq
are complicated non--linear functions of the $F_{\alpha\beta}$ and $\overline{F}_{\dot{\alpha}\dot{\beta}}$ determined via the ``constitutive relations'' \eqref{PfromS}, together with the Bianchi identities
\beq\label{bianchi1}
\partial_{\beta\dot\alpha}\, F_\alpha^\beta \ -\ \partial_{\alpha\dot\beta}\, \bF_{\dot\alpha}^{\dot\beta} \ = \ 0 \ ,
\eeq
are covariant under the duality rotations
\beq
\delta F_{\alpha\beta} \ = \ \eta\, P_{\alpha\beta}\ ,\quad \delta P_{\alpha\beta} \ =\ - \ \eta\, F_{\alpha\beta} \ , \label{FP_duality}
\eeq
in analogy with the free Maxwell system.

The natural mixing of eqs.~\eqref{EoM1} and \eqref{bianchi1} is indeed strikingly compatible with the origin of $P$ and $\overline{P}$ from the BI action via eq.~\eqref{PfromS}. This crucial consistency condition and its multi--field extensions were later formulated systematically by GZ in \cite{GZ} and in \cite{subsequent}. In the single--field case there is a single constraint,
\beq
F^2 \ + \ P^2 \ - \ \bF^2 \ - \ \bP^2 \ = \ 0\ , \label{gz_1}
\eeq
which holds identically, as one can verify, for the BI theory.

The relevance of the tensor auxiliary variables $V_{\alpha\beta}$ and $\overline{V}_{\dot{\alpha}\dot{\beta}}$ goes well beyond the simplifications evident in eq.~\eqref{BI12}. While other options have been explored to linearize the BI action, as in \cite{RT}, the auxiliary fields $V_{\alpha\beta}, \bV_{\dot\alpha\dot\beta}$ possess a special virtue: duality transformations act linearly on them, according to
\beq\label{dual1}
\delta V_{\alpha\beta}\ = \ -i \, \eta\, V_{\alpha\beta}, \qquad \delta \bV_{\dot\alpha\dot\beta} \ = \ i\,\eta\, \bV_{\dot\alpha\dot\beta}\ ,
\eeq
\emph{in a universal fashion} that is independent of the dynamics. These relations, whose origin we are about to review, clearly imply that $a$ in eq.~\eqref{nu_nubar_a} \emph{is invariant under the duality}, and thus retain their form even if $V_{\alpha\beta}$ and $\bV_{\dot\alpha\dot\beta}$ are rescaled by an arbitrary function ``lapse function'' $h(a)$.

The reader should appreciate the sharp contrast between eq.~\eqref{dual1} and the effect of duality transformations on the ordinary variables, since the actual nature of the $P_{\alpha\beta}$ and $\overline{P}_{\dot\alpha\dot\beta}$ reflects the specific form of the Lagrangian. The striking simplification inherent in eq.~\eqref{dual1} makes it possible to address dualities and corresponding generalizations of the BI theory in a systematic fashion.

\subsection{The Master Action} \label{sec:2.1}

In addressing generalized dualities, it is convenient to rely on ``master actions'' that combine the dynamical curvature $F_{\alpha\beta}$ and the auxiliary field $V_{\alpha\beta}$ with their complex conjugates. For the one--field systems of interest in this section, these are built integrating over space time the Lagrangians
\beq\label{L2VF}
{\cal L} \ = \ \frac{1}{2}\,\left(\phi+\overline \phi \right)\ -\ 2\, h\, \left(F\cdot V+\overline F\cdot \overline V\right)\ +\ h^2\, \left(\nu \ + \ \overline \nu \right)+E\big(\nu,\,\overline \nu \big)\ .
\eeq
These rest on generic Lorentz--invariant interaction terms $E\big(\nu,\,{\overline \nu}\big)$, and extend slightly the result of the second paper in \cite{Iv}, since they also involve the duality--invariant scalar ``lapse function'' $h\big(a\big)$, which will prove very useful in the following. The BI action is a special case, and is recovered if
\bea
&& E \ = \ 2\,a\,\frac{1 \ + \ a}{\left(1\ -\ a \right)^2}\ ,\\
&& h \ = \ \frac{\sqrt{2}}{1 \ - \ a}\ . \
\eea
In the following we would like to characterize, following IZ \cite{Iv}, the subset of actions whose equations of motion are invariant under the duality \eqref{gz_1}, where now
\beq
P_{\alpha\beta}(F,V)\ = \ i\,(F_{\alpha\beta}\ -\ 2\,h\, V_{\alpha\beta})\ , \label{def_P}
\eeq
and to display a deformation of the BI example. Let us notice aforehand that, when combined with eq.~\eqref{FP_duality}, this relation implies the universal linear duality transformations for $V_{\alpha\beta}$ and $\overline{V}_{\dot{\alpha}\dot{\beta}}$ of eq.~\eqref{dual1}.

The equations of motion resulting from the
Lagrangian (\ref{L2VF}) and the corresponding Bianchi identities can be duality--covariant only for suitable choices of the interaction
$E\left(\nu, \overline\nu\right)$. The restriction, embodied in the
constraint \eqref{gz_1}, can be recast in a form that makes its
group-theoretical meaning quite transparent.

The equations linking $F_{\alpha\beta}$ to $V_{\alpha\beta}$ and $\overline
V_{\dot\alpha\dot\beta}$ play a key role in the formalism.
They obtain since the Lagrangian ${\cal L}(V, F)$ is to be stationary with respect to variations of the auxiliary fields, and read
\bea \label{FVequ}
F_{\alpha\beta}=\left(h+\frac{2 \,\overline \nu\,\partial_a h \,\left(\nu \, \partial_\nu \,E \ - \ \overline{\nu}\, \partial_{\overline{\nu}} \,E \right)\ +\ h\, \partial_\nu \, E}{h\,\left(4\, a\, \partial_a h \ + \ h\right)}\right)\,V_{\alpha\beta} \qquad
\mbox{(and \ c.c.)}\ .
\eea
They relate, for any dynamical model, $\phi$ and $\overline{\phi}$ to the quantities listed in eq.~\eqref{nu_nubar_a}. Using this result and definition of $P$ in eq.~\eqref{def_P}, one can recast eq.~(\ref{gz_1}) in the form
\beq
\nu \ \partial_\nu \,E \ - \ \overline{\nu}\ \partial_{\overline{\nu}} \,E \ = \ 0 \ .
\eeq
This first--order equation demands that $E$ depend on the auxiliary fields only via the scalar $a$ of eq.~\eqref{nu_nubar_a}, which is clearly invariant under the $U(1)$ duality.

If $E=E(a)$, eq.~\eqref{FVequ} simplifies considerably and reduces to
\beq\label{ing}
F_{\alpha\beta} \ =\ \left(h\ +\ p\, \overline \nu\right)V_{\alpha\beta} \qquad
\mbox{(and \ c.c.)}\ ,
\eeq
where
\beq
p\ = \ \frac{E_a}{4\,a\,\partial_a h \, +\, h}\ , \label{p_onefield}
\eeq
and eq.~\eqref{ing} implies the two useful results
\bea
F\cdot V\ =\ (h\ +\ p\, \overline \nu)\,\nu \qquad
\mbox{(and \ c.c.)}\ ,\\
\phi\ =\ (h\ +\ p\, \overline \nu)^2\,\nu  \label{phi_nu} \qquad
\mbox{(and \ c.c.)}\ .
\eea
In terms of the auxiliary variables, the Lagrangian (\ref{L2VF}) reduces to
\beq\label{Lag1}
{\cal L} \ =\ - \ \frac{1}{2}\,\left( \nu+ \overline{\nu}\right) \left( h^2 \ - \ a\, p^2\right) \ +  \ I(a)\ ,
\eeq
where
\beq
I(a) \ =\ E\ -\ 2\, a\, h\, p \ , \label{I_a}
\eeq
and on account of eq.~\eqref{p_onefield} $I$ is determined by the differential equation
\beq\label{rho}
\partial_a I \ =\ -\ h\, p \ + \ 2\, a \left( p\, \partial_a h \ -\ h\, \partial_a p \right)\ .
\eeq

Any choice of $I(a)$ yields a duality invariant model, but one must eventually return to the standard variables $F_{\alpha\beta}$ and $\overline{F}_{\dot{\alpha}\dot{\beta}}$, and thus, on account of Lorentz invariance, to $\phi$ and $\overline{\phi}$ of eq.~\eqref{phi_phibar}. The relevant information is contained in eq.~\eqref{phi_nu}, but the inversion problem is typically complicated and closed--form expressions for the Lagrangian obtain only in a limited number of cases.

\subsection{Explicit Solutions}\label{sec:2.2}

To begin with, in the weak limit for the interactions
\beq\label{m1}
p\  \simeq \ 0 \ ,
\eeq
and one recovers the Maxwell Lagrangian, which in two--component notation reads
\beq\label{m11}
{\cal L} \ =\ - \ \frac{1}{2}\,\left( F^2 + \bF^2\right)\ .
\eeq
Formally, one might also contemplate the opposite limit
\beq\label{m12}
h \ \simeq \ 0 \ ,
\eeq
which amusingly leads to the Lagrangian
\beq\label{m13}
{\cal L} \ = \ \frac{1}{2}\,\left( F^2 + \bF^2\right) \ ,
\eeq
where the roles of electric and magnetic fields are somehow interchanged.

In general, if both $h$ and $p$ are nonzero, it proves convenient to choose the \emph{gauge}
\beq\label{n1}
h\ =\ p \ .
\eeq
The important step, as we have stated already, is to find $a$ in terms $\phi$ and $\overline \phi$, and to this end let us note the two consequences of eq.~\eqref{phi_nu},
\bea\label{neq12}
&&\nu \ + \ \overline{\nu} \ = \ \frac{\phi\ + \ \overline{\phi} \ - \ 4\, a\, h^2(a)}{h^2(a)(1 \ + \ a)} \ , \label{neq22} \\
&&\phi \,\overline{\phi}\left(1\ + \ a\right)^2 \ = \ a \left[h^2(a) \left(1\ - \ a\right)^2 \ + \ \phi\ + \ \overline{\phi} \right]^2  \ . \label{eq_a_1fields}
\eea
Making use of eq.~\eqref{neq22}, the Lagrangian can be recast in the form
\beq
{\cal L} \ =\ - \ \frac{1}{2}\,\left( \phi+ \overline{\phi}\right) \frac{ 1 \ - \ a\, }{1\ + \ a}\ + \ {2\,a\,h^2}\ \frac{ 1 \ - \ a\, }{1\ + \ a} \ +  \ I(a)\ , \label{lag_reduced_onefield}
\eeq
where
\beq
\partial_a\, I \ =\ -\ h^2(a) \ .
\eeq
The transition to the final form in terms of space--time fields rests on the elimination of $a$ via eq.~\eqref{eq_a_1fields}, which is simple only for special choices of the ``lapse function'' $h(a)$, and thus of the interaction terms $I(a)$ or $E(a)$.

The family of choices
\beq\label{nf}
h^2\ =\ \frac{\beta+\alpha\, \sqrt{a}+\gamma\, a}{\sqrt{a}\,(1-a)^2} \ ,
\eeq
with $\alpha$, $\beta$ and $\gamma$ three constants leads to simple solutions of eq.~\eqref{eq_a_1fields} for $a$. Indeed, while it would turn eq.~\eqref{eq_a_1fields} into a fourth--order equations, the latter is the perfect square of
\beq
\sqrt{\phi \,\overline{\phi}} \left(1\ + \ a\right) \ = \ \sqrt{a} \left[h^2(a) \left(1\ - \ a\right)^2 \ + \ \phi\ + \ \overline{\phi} \right]  \ ,
\eeq
which can be easily solved for all this choices, with the end result
\beq\label{nrho}
I\ =\ \delta\ -\ \frac{\alpha+(\beta+\gamma)\,\sqrt{a}}{1-a}\ -\ (\beta-\gamma)\, \mbox{ ArcTanh } \left(\sqrt{a}\right) \ .
\eeq
In terms of auxiliary variables, the corresponding Lagrangians read
\beq\label{nLag}
{\cal L} \ =\ -\ \frac{1}{2}\,\frac{\beta+\alpha \sqrt{a}+\gamma a}{\sqrt{a}\,(1-a)}\,\left( \nu+ \overline{\nu} \right) \ + \ I \ ,
\eeq
and the appropriate Maxwell limit obtains provided one chooses $\beta=0$, $\alpha=2$ and the integration constant $\delta=2$. Doing this and solving eq.~\eqref{neq22} for $a$ yields
\bea
{\cal L} &=& f^2\left[1\ -\ \sqrt{\left(1+\frac{F^2+\overline{F}^2}{2\, f^2}\right)^2\,-\, \frac{1}{f^2}\,\sqrt{F^2\, \bF^2}\,\left(\frac{1}{f^2}\,\sqrt{F^2\, \bF^2}\,-\,\gamma\right)} \right.\label{c}\\
&+&\ \left. \gamma\,
\mbox{ArcTanh} \left(\frac{1+\frac{F^2+\overline{F}^2}{2\,f^2}-\sqrt{\left(1+\frac{F^2+\overline{F}^2}{2\,f^2}\right)^2\,-\, \frac{1}{f^2}\,\sqrt{F^2\, \bF^2}\,\left(\frac{1}{f^2}\,\sqrt{F^2\, \bF^2}\,-\,\gamma\right)}}{\frac{1}{f^2}\ \sqrt{F^2\, \bF^2}-\gamma}\right)\right]\ , \nonumber
\eea
where we have also reinstated the scale $f$ of eq.~\eqref{BI}.
Notice that these models involve the combination $\sqrt{F^2\, \bF^2}$, which is not analytic at the origin of field space. Still, one can argue on the basis of standard theorems of Calculus that their behavior is regular enough to grant a well--defined Cauchy problem. This type of feature will show up again in the following section. The choice $\gamma=0$ clearly recovers the standard BI action, whose form in auxiliary variables was already given in \eqref{BI12}.

\section{Two--field Models with Extended Dualities}\label{sec:3}

We can now move on to a less explored territory. Our next aim is to construct examples of non--linear Lagrangians for a pair of field strengths  $F^i_{\alpha\beta},\,\overline{F}^i_{\dot\alpha\dot\beta}$,\ $(i = 1,\,2)$. As we have anticipated, we shall rely on a slight generalization of the approach spelled out in the last paper in \cite{Iv}, which will rest again on a ``lapse function'' $h$. Our main result will be a new explicit solution with $U(2)$ duality, but the same techniques will also recover, in a clear fashion, other models that we had recently obtained less systematically in \cite{FSY}, with $SU(2)$ and $U(1) \times U(1)$ duality groups. On the other hand, the model in eq.~(3.13) of \cite{FSY} with manifest $U(1)$ symmetry does not belong to this list, despite its double self--duality under Legendre transforms of both $F$ and $G$. It lacks in fact the simultaneous presence of electric and magnetic duality generators, which is instrumental in making the IZ method particularly effective.

We shall restrict again our attention to Lagrangians
\beq
{\cal L}(F^k,\,\overline{F}^l) \qquad \left(k,l = 1,\,2\right)
\eeq
that are manifestly invariant under Lorentz transformations and under the $O(2)$
transformation
\beq
\delta_{\xi} F^k_{\alpha\beta}\ =\ \xi^{kl}\,F^l_{\alpha\beta}\ ,\quad
\delta_{\xi}\overline{F}^k_{\dot\alpha\dot\beta}\ =\ \xi^{kl}\,\overline{F}^k_{\dot\alpha\dot\beta}\,,\quad
\xi^{kl}=-\xi^{lk}\ .\label{ON}
\eeq

As in the previous section (see eq.~\eqref{phi_phibar}), Lorentz invariance leads one to define complex scalar variables, which are now the matrices
\beq
\phi^{kl}\ = \ F^k \cdot F^l\ ,\qquad
\overline\phi^{kl} \ = \ \overline{F}^k \cdot \overline{F}^l \ ,
\eeq
and to regard the Lagrangian as a real function of them. The resulting non--linear equations of motion
\bea
&&E^k_{\alpha\dot\alpha}\ \equiv \ \partial_\alpha^{\dot\beta}\
\overline{P}^k_{\dot\alpha\dot\beta}(F) \ - \ \partial^\beta_{\dot\alpha}\
P^k_{\alpha\beta}(F)\ = \ 0 \ \label{eomF}
\eea
involve the dual nonlinear field strengths
\bea P^k_{\alpha\beta}(F)\ \equiv \ i\,\frac{\partial L}{\partial
F^{k\alpha\beta}} \ = \ 2\,i\,F^l_{\alpha\beta}\,\frac{\partial L}{\partial
\varphi^{kl}} \qquad \mbox{(and\ c.c.)}\ ,
\eea
while the ordinary field strengths $F^{k}_{\alpha\beta}$, $\overline F^{k}_{\dot\alpha\dot\beta}$ obey the Bianchi identities
\bea
&&B^k_{\alpha\dot\alpha}\ \equiv \ \partial_\alpha^{\dot\beta}
\overline{F}^k_{\dot\alpha\dot\beta} \ - \ \partial^\beta_{\dot\alpha}
F^k_{\alpha\beta} \ = \ 0\ . \label{Bianchi}
\eea

As in the last paper in \cite{Iv}, the master actions for these manifestly $U(1)$ invariant Lagrangians rest on complex auxiliary tensor fields $V^k_{\alpha\beta}\,$, $\overline V^k_{\dot\alpha\dot\beta}\,$. However, here they also depend on a ``lapse function'' $h$, and read
\bea
{\cal L}&=&\frac12\,\left(\phi_t+\overline \phi_t\right)\ -\ 2 \,h\,\left(F^k\cdot V^k+\overline{F}^k\cdot \overline{V}^k\right)\ +\ h^2\,\left(\nu_t+\overline \nu_t\right)\ +  \ E\big(\nu^{kl},\,\overline\nu^{kl} \big)\ ,
\label{LFVN}
\eea
where
\beq
\nu^{kl}\ = \ V^k\cdot V^l\ , \quad \nu_t \\ = \ {\rm Tr}\left(\nu\right) \ , \qquad \mbox{(and \ c.c.)} \ .
\eeq
From now on, the suffix $t$ will identify, for brevity, a \emph{trace} of the corresponding matrix.
Eq.~\eqref{LFVN} implies that
\beq\label{pk}
P_{\alpha\beta}^k\ = \ i\, \left(F_{\alpha\beta}^k\ -\ 2 \,h\, V_{\alpha\beta}^k\right)\ ,
\eeq
and in all these constructions the function $h$ {will be invariant under the full duality at stake}.

The algebraic equations for $V^k_{\alpha\beta}, \overline
V^k_{\dot\alpha\dot\beta}$ obtain varying ${\cal L}(V, F)$ with respect to
$V^k_{\alpha\beta}, \, \overline V^k_{\dot\alpha\dot\beta}$, define the ordinary field strengths in terms of the auxiliary tensors, and are of the form
\beq\label{fk}
F^k \ = \ h\, V^k\ + \ g^{kn}\,V^n\ , \qquad
\mbox{(and \ c.c.)}\ ,
\eeq
where
\beq\label{gmxdef}
g^{kn} \ = \ \frac{1}{h}\,\left[\frac{\partial E}{\partial \nu^{kn}}\ -\ R \,\frac{\partial h}{\partial \nu^{kn}}\right] \ ,
\eeq
with
\beq
R \ = \  \frac{\nu^{ml} \frac{\partial E}{\partial \nu^{ml}}\ +\ \overline \nu^{ml} \frac{\partial E}{\partial \overline \nu^{ml}}}{\nu^{ml} \frac{\partial h}{\partial \nu^{ml}}\ +\ \overline \nu^{ml} \frac{\partial h}{\partial \overline \nu^{ml}}\ +\ \frac{1}{2}\, h} \ . \label{Rdef}
\eeq

Depending on the actual duality symmetry, the following ``magnetic'' GZ constraints
\bea
{\cal M}^{kl}\ \equiv \ (P^kP^l)\ +\ (F^kF^l)\ -\  \mbox{c.c.}\ = \ 0 \  ,
\eea
or at least some combinations thereof,
will hold. On the other hand, the ``electric'' GZ constraint
\beq\label{elec}
{\cal E}^{kl}\ \equiv \ (F^kP^l) \ - \ (F^l P^k)\ - \ \mbox{c.c.} \ = \ 0 \ ,
\eeq
which is unique in the two--field case, will always hold as a result of the manifest $U(1)$ symmetry that we have assumed for the Lagrangians. In detail, this $U(1)$ symmetry means that all terms in the Lagrangian can only depend, a priori, on the five independent variables
\beq\label{5var}
\nu_t \ \equiv \ {\rm Tr}\left(\nu\right) \ ,\quad \overline\nu_t\ \equiv \ {\rm Tr}\left(\overline \nu\right) ,\quad a_t\ \equiv \ {\rm Tr}\left({\cal A} \right)\ , \quad \nu_d\ \equiv \ {\rm Det}\left(\nu\right) ,\quad \overline\nu_d \ \equiv \ {\rm Det}\left(\overline\nu\right) \ ,
\eeq
where the Hermitian matrix ${\cal A}$ is defined as the product of the two matrices $\overline \nu$ and $\nu$:
\beq
{\cal A} \ = \overline \nu \, \nu \ , \qquad a_d \ = \  \mbox{Det} \left( {\cal A} \right) \ . \label{anunubar}
\eeq
Clearly the determinant of ${\cal A}$, which we shall call $a_d$ in the following, is not an independent quantity. Rather, it is simply the product of $\nu_d$ and $\overline{\nu}_d$.

These results can be understood as follows. To begin with, eqs.~\eqref{ON}, \eqref{pk} and \eqref{fk} imply that the ``electric'' $U(1)$ transformations within $U(2)$ act on $V^k_{\alpha\beta}\,$ and $\overline V^k_{\dot\alpha\dot\beta}\,$  according to
\beq
\delta V^k_{\alpha\beta}\ =\ \xi^{kl}\,V^l_{\alpha\beta}\qquad \mbox{(and c.c.)} \ ,
\eeq
and therefore
\beq
\delta\nu\ =\ [\xi,\nu] \ ,
\eeq
so that $\nu_t$, $\nu_d$ and $a_t$ and, a fortiori, $a_d$, are all invariant under the ``electric'' $U(1)$.

Making use of eqs.~\eqref{pk} and \eqref{fk}, the GZ constraints take the form
\bea
&&{\cal M}^{kl} \ \equiv \ \left(g^{kn}\,\nu^{nl} \ + \ g^{ln}\,\nu^{nk}\right) \ - \ {\rm c.c.}\ = \ 0 \ ,\label{eqgM}\\
&&{\cal E}^{kl} \ \equiv \ \left(g^{kn}\,\nu^{nl} \ - \ g^{ln}\,\nu^{nk} \right)\ + \ {\rm c.c.}\ = \ 0 \ .\label{eqgE}
\eea
Notice also that, on account of the manifest $U(1)$ ``electric'' duality symmetry, the matrix $g$ reduces to
\beq
g^{kn} \ = \ p\ \overline \nu^{kn} \ +\ q\,\left(\nu^{-1}\right)^{\,kn}\ + \ r\ \delta^{kn}\ ,
\eeq
where $p$ is a real function while $q=q_1+i\, q_2$ and $r=r_1+i\, r_2$ are complex functions, all built out of derivatives of the ``interaction'' term $E$ and of the ``lapse function'' $h$ with respect to the five invariants of eq.~\eqref{5var}. In detail:
\beq
p \ = \ \frac{1}{h} \left[\frac{\partial E}{\partial {a_t}}\ -\ R \,\frac{\partial h}{\partial {a_t}}\right] \, ,\ \
q \ = \ \frac{\nu_d}{h} \left[\frac{\partial E}{\partial {\nu_d}}\ -\ R \,\frac{\partial h}{\partial {\nu_d}}\right] \, , \ \
r \ = \ \frac{1}{h} \left[\frac{\partial E}{\partial {\nu_t}}\ -\ R \,\frac{\partial h}{\partial {\nu_t}}\right] \ . \label{defpqr}
\eeq

At this point, the ``electric'' GZ \eqref{eqgE} constraint is identically satisfied while the three ``magnetic'' GZ constraints \eqref{eqgM} can be cast in the convenient form
\bea
{\cal M}^{12} & \sim & r_1\,\Im[\nu^{12}]\ +\ r_2\,\Re[\nu^{12}]\ ,\label{12}\label{min1}\\
{\cal M}^{11} \ + \ {\cal M}^{22}&\sim & 2\,q_2\ + \ r_1\,\Im[\nu_t]\ +\ r_2\,\Re[\nu_t]\ ,\label{tr}\label{min2}\\
{\cal M}^{11} \ - \ {\cal M}^{22}& \sim & r_1\,\Im[\nu^{11}-\nu^{22}]\ +\ r_2\,\Re[\nu^{11}-\nu^{22}]\,\label{min3} \ ,
\eea
where the second of these equations only involves invariants of the ``electric'' $U(1)$ duality group. More in detail, the second constraint corresponds to the $U(1)$ generator in $U(2)$ that commutes with all others, while the first and third constraints correspond to the two generators that close, together with the ``electric'' generator, into the $SU(2)$ algebra.

These equations merely identify the types of the solutions, which fall into three classes associated with $U(2)$, $SU(2)$ and $U(1) \times U(1)$ duality symmetry. Arriving at explicit examples entails a main complication, the inversion problem to recover their forms in terms of standard variables.

\subsection{A Model with $U(2)$ Duality}\label{sec:3.2}

In order to attain $U(2)$ duality, all three equations of the system \eqref{min1}--\eqref{min3} must be satisfied for generic values of the $\nu^{ij}$. Thus, $r_1$, $r_2$ and $q_2$ must vanish, and these conditions imply that $h$ and $E$ can only depend on $a_t$, the trace of ${\cal A}$, and on its determinant $a_d$.

One can also state, equivalently, that two--field models admitting the maximal $U(2)$ duality symmetry must be also compatible with three ``magnetic'' transformations realized as
\beq
\delta_{\eta}
F^k_{\alpha\beta}\ =\ \eta^{kl}P^l_{\alpha\beta}\ , \quad \delta_{\eta} P^k_{\alpha\beta}\ =\ -\ \eta^{kl}F^l_{\alpha\beta}\ . \label{uN1}
\eeq
Here the symmetric matrix $\eta^{kl}$ encodes three real parameters, and the equations of motion \p{eomF} and the Bianchi identities \p{Bianchi} are to be covariant under eq.~\eqref{uN1}. The $U(2)$ transformations for the auxiliary tensor fields $V^k_{\alpha\beta}\,$ and $\overline V^k_{\dot\alpha\dot\beta}\,$ read
\bea
&&\delta V^k_{\alpha\beta}\ =\ \left(\xi^{kl}\ -\ i\,\eta^{kl}\right)V^l_{\alpha\beta}\ ,
\quad
\delta\overline{V}^k_{\dot\alpha\dot\beta}\ =\ \left(\xi^{kl}\ +\ i\,\eta^{kl}\right)
\overline{V}^l_{\dot\alpha\dot\beta}\ ,
\eea
where the antisymmetric matrix associated to the ``electric'' $U(1)$ was introduced in eq.~\eqref{ON}.
These transformations imply corresponding ones for the complex scalar variables $\nu^{kl},\,\overline \nu^{kl}$, which can be summarized in the compact matrix form
\bea
\delta\nu=[\xi,\nu]-i\{\eta,\nu\}\,,\quad
\delta\overline\nu=[\xi,\overline\nu]+i\{\eta,\overline\nu\}\,. \eea
Consequently, the Hermitian matrix ${\cal A}$ transforms as
\bea
\delta {\cal A}\ =\ [\xi\ +\ i\,\eta,{\cal A}] \ ,
\eea
and one can indeed recover the two $U(2)$ invariants that we had identified starting from eqs.~\eqref{min1}--\eqref{min3}, the trace $a_t$ of ${\cal A}$ and its determinant $a_d$.

Eqs.~(\ref{min1})--(\ref{min3}) all vanish for this class of models, and as we have explained $r=0$ and $q$ is purely real and equal to $q_1$. As a result, the field strengths $F^k_{\alpha\beta}$ and their duals $P^k_{\alpha\beta}$ can be represented as
\bea
F^k_{\alpha\beta}&=& \left(h\, \delta^{kl}\ +\ p\,\overline{\nu}^{kl}\ +\ q_1\, \nu^{-1\,kl}\right)\,V^l_{\alpha\beta} \ , \\
P^k_{\alpha\beta}&=&\left(-\,h\, \delta^{kl}\ +\ p\,\overline{\nu}^{kl}\ +\ q_1\, \nu^{-1\,kl}\right)\,V^l_{\alpha\beta}\ .
\eea
where $p$ and $q_1$ take the form
\bea
p&=&\frac{1}{h}\,\frac{h\, {\partial_{a_t}} E \ +\ 8\, a_d\,\left({\partial_{a_d}}h\, {\partial_{a_t}}E \ - \ {\partial_{a_t}}h\,{\partial_{a_d}}E\right)}{4\, a_t\, {\partial_{a_t}} h \ +\ 8\, a_d \, {\partial_{a_d}}h \ +\ h}\ ,\\
q_1&=& \frac{1}{h}\,\frac{h\, a_d\, {\partial_{a_d}}E \ -\ 4\, a_d\,a_t\,\left({\partial_{a_d}}h\, {\partial_{a_t}}E \ - \ {\partial_{a_t}}h\,{\partial_{a_d}}E\right)}{4\, a_t\, {\partial_{a_t}} h\ +\ 8\, a_d \, {\partial_{a_d}}h \ +\ h}\ . \label{q1_reduced}
\eea
Notice that $q_1$ is a key new ingredient, which had no analogue in the one--field case. One is thus led to Lagrangians of the form
\bea\label{lagu21}
{\cal L} \ =\ \left[ \frac12\,\left(-h\, \delta^{kn}+p\,\overline{\nu}^{kn}+q_1\, \nu^{-1\,kn}\right)\,\left(h\, \delta^{nl}+p\,\overline{\nu}^{nl}+q_1\, \nu^{-1\,nl}\right)^{-1}\varphi^{lk}+\mbox{c.c.}\right] +I\ ,
\eea
where
\beq
I \ =\ E\ -\ 2\, a_t\,p\,h -\ 4\,q_1\,h
\eeq
is to satisfy the two conditions
\bea
\partial_{a_t} I&=&-\ h \,p\ +\ 2\,a_t \left(p\, \partial_{a_t} h \ - \ h\,\partial_{a_t} p\right)\ +\ 4\, \left(q_1\, \partial_{a_t} h\, \ - \ h \, \partial_{a_t} q_1 \right)\ ,\\
\partial_{a_d} I &=&\frac{h \,q_1}{a_d}\ +\ 2\, a_t\, \left( p\, \partial_{a_d} h \ - \ h\,\partial_{a_d} p \right)\ +\ 4\, \left( q_1\, \partial_{a_d} h\, \ - \ h \, \partial_{a_d} q_1 \right)\,.
\eea
In analogy with the one--field case, it is convenient to regard $p$ and $q_1$ as independent variables, but here one is also to verify the integrability condition
\bea
\label{int}
a_d \,\partial_{a_d} h \left[3\, p\,+\,4\left( a_t\, \partial_{a_t} p \,+\,2\, \partial_{a_t} q_1\right)\right]&+& \partial_{a_t}h\, \left[q_1\,-\,4\, a_d\,(a_t \,\partial_{a_d} p\,+\,2\,\partial_{a_d} q_1)\right]\nonumber \\ &+& h\,\left(\partial_{a_t}q_1\,-\,a_d \,\partial_{a_d} p\right) \ =\ 0\ .
\eea

One can recast the Lagrangian in a form that only involves the auxiliary variables,
\bea\label{lag2inv}
{\cal L} \ =\ \frac12\left(-h^2+2\, p\,q_1+a_t\, p^2\right)\, (\nu_t+\overline{\nu}_t)\ +\ \frac12\,\left(-p^2+\frac{q_1^2}{a_d}\right)\,\left(\nu_t\,\overline{\nu}_d+\overline{\nu}_t\,\nu_d\right)\ +\ I\ ,
\eea
but the eventual conversion of ${\cal L}$ into normal variables rests on the possibility of solving the algebraic equations
\bea\label{defphi}
\phi^{kl}=\left(h\, \delta^{kn}\ +\ p\,\overline{\nu}^{kn}\ +\ q_1\, \nu^{-1\,kn}\right)\,\nu^{ns}\,\left(h\, \delta^{sl}\ +\ p\,\overline{\nu}^{sl}\ +\ q_1\, \nu^{-1\,sl}\right)
\eea
and their complex conjugates for the five variables of eq.~\eqref{5var}.

So far we have been completely general, but our aim is to provide some instructive examples, and one can see that the Lagrangian \eqref{lag2inv} simplifies drastically if
\beq
q_1\ =\ \sqrt{a_d}\,p \ . \label{conj}
\eeq

Choosing, as in the one--field case, the gauge $h=p$, the self--consistency condition (\ref{int}) reduces to %
\beq
\sqrt{a_d}\ \partial_{a_d} h \ -\ \partial_{a_t} h\ =\ 0\ ,
\eeq
which is simply solved provided $h$ depends on $a_t$ and $a_d$ only via the combination
\beq
a\ =\ a_t+2\,\sqrt{a_d} \ . \label{combination}
\eeq
Let us stress that the solution considered in \cite{ABMZ} does not belong to this class. We shall return to this point shortly.

All in all, in this fashion the Lagrangian (\ref{lagu21}) reduces to
\beq\label{lagu22}
{\cal L} \ =\ - \ \left(1\,-\,a\right)h(a)^2\,Re [\nu_t]\ +\ I(a) \ ,
\eeq
with
\beq
\quad \partial_a I \ =\ -\ h^2\ ,
\eeq
in striking analogy with eq.~\eqref{lag_reduced_onefield} for the one--field case.

Using the definition (\ref{defphi}) of the matrix $\phi$ in terms of the auxiliary variables, one can set up the inversion problem to ordinary field variables via the following relations:
\bea
\phi_t &=& h^2 \left[ \nu_t \ + \ \ a( \overline{\nu}_t \ + 2) \right] \ , \label{nueq}\\
{\rm Det}(\phi - \overline{\phi}) &=& h^4 \left(1
\ - \ a\right)^2 \left(\nu_d \ + \overline{\nu}_d \ + \ a \ - \ \nu_t \, \overline{\nu}_t \ - \ 2\, \sqrt{a_d}\right) \nonumber \\
&=& h^4 \left(1 \ - \ a\right)^2
\left[(\sqrt{\nu_d} \ - \sqrt{\overline{\nu}_d})^2 \ + \ a \ - \ \nu_t \, \overline{\nu}_t \right] \label{aeq} \ , \\
\phi_d &=& \frac{h^4}{\nu_d} \left[ \nu_d(1\ + \ \overline{\nu}_t)
 + \ \sqrt{a_d} (\nu_t \ + \ a)\right]^2 \nonumber \\
&=& h^4 \left[ \sqrt{\nu_d}(1\ + \ \overline{\nu}_t)
 + \ \sqrt{{\overline{\nu}}_d} (\nu_t \ + \ a)\right]^2 \ . \label{nudeq}
\eea
Using these expressions, one thus arrives at the important equation
\beq
a\,\left(h_1^2+2 \,\Re[\phi_t]\right)^2 \ - \ \left(1+a\right)^2\,\left[ {\rm Det}[\phi-\overline \phi]+\left|\phi_t\right|^2-2\left(\Re[\phi_d]\,-\, \sqrt{\left|\phi_d\right|^2}\right) \right]=\ 0\ ,\label{nequ2}
\eeq
where
\beq
h_1\ =\ \left(1\ -\ a\right)\,h\ ,
\eeq
which is the counterpart of eq.~\eqref{eq_a_1fields} of the one--field case. Notice however the presence of the square root in the last term, which brings this construction beyond the framework considered by \cite{ABMZ}, and the implicit positivity condition on the last group of terms, which will be important for the final Lagrangian that we are about to display.

The simplest choice for $h_1$ that makes it possible to solve eq.(\ref{nequ2}) analytically is
\beq\label{condf1}
h_1\ =\ \sqrt{2} \  \ \longrightarrow \ \ h \ = \ \frac{\sqrt{2}}{1 \ - \ a} \ .
\eeq
In this case eq.~\eqref{nequ2} becomes quadratic, and the Lagrangian (\ref{lagu22}) takes again the form that we already came across in eq.~\eqref{BI12},
\beq\label{lagu2ax}
{\cal L}\ =\ -\ \frac{2}{1\ -\ a}\,\bigl(\Re[\nu_t]\ +\ a\bigr)\ ,
\eeq
where the choice in eq.~\eqref{condf1} also guarantees the correct weak--field limit.
This Lagrangian is formally identical to the one previously considered in the last paper in \cite{Iv} with reference to the construction in \cite{ABMZ}, but for a crucial difference. We started from the condition \eqref{conj}, which was motivated by the simplifications it brought about and led to identify the combination $a$ of eq.~\eqref{combination}. On the other hand, the authors of \cite{ABMZ} demanded that there be no dependence on $a_d$, which led to the identification of $a$ with $a_t$ and to the condition that $q_1$ vanish, as can be seen from eq.~\eqref{q1_reduced}. All in all, it was then impossible, in \cite{ABMZ}, to perform the inversion analytically.

With our choices one can now revert to the ordinary variables $\phi^{kl}$, solving eq.~(\ref{nequ2}) for $a$ with $h_1$ as in (\ref{condf1}) and substituting in the Lagrangian (\ref{lagu2ax}). The end result (with the scale $f$ of eq.~\eqref{BI} set to one for brevity),
\beq
{\cal L}\ =\ 1\ -\ \sqrt{\,\left(1+\Re[\phi_t]\right)^2-\left|\phi_t\right|^2-{\rm Det}[\phi- \overline{\phi}]+2\,\left(\Re[\phi_d]\ - \ \sqrt{\left|\phi_d\right|^2}\right)} \ , \label{lagfinu2}
\eeq
has $U(2)$ duality and reduces to the BI theory if the two Abelian field strengths coincide. Notice the peculiar inner square root, whose argument is positive semi--definite but is not analytic at the origin of field space. Notice also that, on account of eq.~\eqref{nequ2}, the combination of the last four terms inside the outer square root is bound to be \emph{negative}, in analogy with the standard BI case, which is recovered if the two fields are identified.

\subsection{A Model with $SU(2)$ Duality}\label{sec:3.3}

In models with $SU(2)$ duality, only eqs.~\eqref{min1} and \eqref{min3} must be satisfied. This requires, in general, the vanishing of $r_1$ and $r_2$, but not anymore the vanishing of $q_2$. Alternatively, the $\eta$ matrix in the transformations of eq.~\eqref{uN1} is now traceless, and one can see that the remaining conditions imply that $f$ and $E$ can now depend on $a_t$ and on the two combinations $\Re[\nu_d]$ and $\Im[{\nu}_d]$.
As a result, in the $SU(2)$ case the field strength $F^k_{\alpha\beta}$ can still be represented as
\bea
F^k_{\alpha\beta}\ =\ \left(h\, \delta^{kl}\ + \ p\,\overline{\nu}^{\ kl}\ +\ q\, \nu^{-1\ kl}\right)\,V^l_{\alpha\beta}\ ,
\eea
but now $q$ is complex.

In general, in auxiliary variables one is confronted with expressions of the form
\bea
{\cal L}&=&\left[-h^2\,+\,p^2\,a_t\,+\,2\,p\,q_1\,-\,\Re[\nu_d]\,\left(p^2\,-\,\frac{q_1^2-q_2^2}{a_d}\right)\,+\,2\,
\Im[\nu_d]\,\frac{q_1\,q_2}{a_d}\right]\,\Re[\nu_t] \nonumber\\
&+&\left[2 p\, q_2 \,-\,\Im[\nu_d]\,\left(p^2\,-\,\frac{q_1^2\,-\,q_2^2}{a_d}\right)\,-\,2\,\Re[\nu_d]\,\frac{q_1\,q_2}{a_d}\right]\,\Im[\nu_t]\ +\ I\ ,\label{lagsu21}\\
I&=&E\ -\ 2\,a_t\,h\,p\ -\ 4\,h\,q_1\ ,
\eea
but in analogy with what we did in Section \ref{sec:3.2} we shall again restrict our attention to a subclass of Lagrangians that are relatively simple, since they do not depend explicitly on $\Im[\nu_t]$. This condition leads to a quadratic equation for $q_2$, whose solutions are
\beq
q_2\ =\ \frac{\pm\,\sqrt{\Re[\nu_d]^2+\Im[\nu_d]^2}\ +\ \Re[\nu_d]}{\Im[\nu_d]}\,\left(\pm\, p\,\sqrt{\Re[\nu_d]^2+\Im[\nu_d]^2} \ -\ q_1\right)\ . \label{q2_su2}
\eeq
Moreover, ratios disappear if one restricts the attention to a particular choice for $q_1$,
\beq
q_1\ =\ \Re[\nu_d]\,p\ .
\eeq
Indeed, in this case eq.~\eqref{q2_su2} reduces to
\beq
q_2\ =\ \Im[\nu_d]\,p\ ,
\eeq
and the sign choice in it becomes immaterial.

Working again in the gauge $h=p$ one ends up, once more, with the Lagrangian in terms of auxiliary variables of eq.~\eqref{lagu22},
\bea
{\cal L}\ =\ -\ \left(1\ -\ a\right)h(a)^2 \, \Re[\nu_t] \ + \ I(a)\,,\quad \partial_a I\ =\ -\ h^2\ .
\eea
Now, however, $a$ is the combination of $SU(2)$ invariants
\beq\label{asu2}
a\ =\ a_t\ +\ 2\,\Re[\nu_d]\ ,
\eeq
and taking, as in previous section,
\beq
h\ =\ \frac{\sqrt{2}}{1\ -\ a}\ ,
\eeq
the end result is again eq.~(\ref{lagu2ax}) for the Lagrangian in terms of auxiliary variables. To reiterate, the key difference between the $U(2)$ and $SU(2)$ examples that we are presenting lies in the definition of $a$: in Section \ref{sec:3.2} it was the $U(2)$--invariant variable of eq.~\eqref{combination}, while here it is the $SU(2)$--invariant one of eq.~(\ref{asu2}).

Reverting to the field strengths, the Lagrangian takes finally the form
\beq
{\cal L} \ =\ 1\ -\ \sqrt{\,\left(1+\Re[\phi_t]\right)^2-\left|\phi_t\right|^2-{\rm Det}[\phi- \overline{\phi}]}\ . \label{lagfinsu2}
\eeq
Notice how, in this two--field generalization of the BI theory with $SU(2)$ duality, which also reduces to it if the two Abelian field strengths coincide, the square root simply lacks the last contribution present in eq.~\eqref{lagfinu2}. This model was recently discussed in \cite{FSY}, where we obtained it making a peculiar choice for the quartic terms.

\subsection{A Model with $U(1)\times U(1)$ Duality}\label{sec:3.4}

We can now turn to retrieve a Lagrangian with $U(1)\times U(1)$ duality. In this case only the ``magnetic'' GZ equation \eqref{min2} is to be satisfied, together with the ``electric'' one that we enforced to begin with.
Once more, our aim is displaying an example where the inversion problem can be solved in closed form. To this end, a further simplification obtains setting $q_2$ zero, which leads to the constraint
\beq
r_1\,\Im[\nu_t]\ =\ -\ r_2\,\Re[\nu_t]\ .
\eeq
Solving it while taking into account the definitions \eqref{defpqr}, one ends up with a neat result: with this choice the ``interaction'' function $E$ and the ``lapse function'' $h$ depend only on $a_d$,  $a_t$ and $\overline\nu_t\,\nu_t$. As a further simplification, we shall assume that the expressions be also independent of $a_t$ and $a_d$, which automatically implies the vanishing of $p$ and $q_1$. Choosing the gauge $h=r_1$, again with
\beq
h\ = \ \frac{\sqrt{2}}{1-a}\ ,
\eeq
where now
\beq
a\ =\ \overline\nu_t\,\nu_t\ , \label{au1u1}
\eeq
one ends up, once more, with the Lagrangian \eqref{lagu2ax} in terms of auxiliary fields. The difference with respect to the preceding examples originates, once more, from the particular choice of $a$ variable, now given in eq.~\eqref{au1u1}.

In terms of the field strengths, the Lagrangian becomes
\beq
{\cal L} \ =\ 1\ - \ \sqrt{\,\left(1+\Re[\phi_t]\right)^2-\left|\phi_t\right|^2}\ . \label{lagfinu1u1}
\eeq
This is a two--field generalization of the BI theory with $U(1) \times U(1)$ duality, and reduces to it if the two Abelian field strengths coincide. This model was also recently discussed in \cite{FSY}.

\section{Supersymmetry}\label{sec:susy}

The construction that we have illustrated was driven by a search of simple examples realizing the duality groups that are possible with two field strengths. We thus made some choices along the way, which were aimed at attaining handy analytic forms in the inversion. One may wonder whether the explicit Lagrangians that we have built afford a supersymmetric extension. There is a convenient necessary (but not sufficient) condition for ${\cal N}=1$ supersymmetry in multi--field Lagrangians depending on chiral field strengths $W^i_\alpha \equiv \overline{D}^2\,D_\alpha\,V$ and their conjugates. This condition was spelled out in \cite{CF}: in a supersymmetric extension, the quartic terms must be of the form
\beq
I_4 \ = \ \int d^4 \theta \ C_{ijkl} \,  W^{\alpha i}\,W_{\alpha}^j\ \overline{W}^{\dot{\alpha} k}\,\overline{W}_{\dot{\alpha}}^l \ , \label{susy1}
\eeq
and this expression is the supersymmetric completion of
\beq
I_4^B  \ = \ \int d^4 \theta \ C_{ijkl} \,  \left(F_D^{2}\right)^{ij}\, \left(F_A^{2}\right)^{kl} \ , \label{susy2}
\eeq
where the suffixes $D$ and $A$ identify (anti)self-dual combinations. In the two--component notation of the preceding sections, these originate from $F^i_{\alpha\beta}$ (or $\overline{F}^i_{\dot{\alpha}\dot{\beta}}$).  One can now verify whether the quartic terms in eqs.~\eqref{lagfinu2}, \eqref{lagfinsu2} and \eqref{lagfinu1u1} are of this form.

In the $N=2$ case, it is convenient to introduce complex combinations of the two field strengths (here in two--component notation),
\beq
F^{\pm} \ = \ F^1 \ \pm \, i\, F^2 \ ,
\eeq
or of the corresponding ${\cal F}^1$ and ${\cal F}^2$ in four--component notation, and then with a manifest ``electric'' $U(1)$ there are three possible quartic terms,
\bea
&& I_4^{++--}  \ = \ \left(F_D^{2}\right)^{++}\, \left(F_A^{2}\right)^{--} \ , \\
&& I_4^{--++}  \ = \ \left(F_D^{2}\right)^{--}\, \left(F_A^{2}\right)^{++} \ , \\
&& I_4^{+-+-}  \ = \ \left(F_D^{2}\right)^{+-}\, \left(F_A^{2}\right)^{-+} \ .
\eea
Notice that all these invariants are real, since
\beq
\left( F_D^+ \right)^\star \ = \ \left( F_A^- \right) \ , \qquad \left( F_D^- \right)^\star \ = \ \left( F_A^+ \right) \ .
 \eeq

Making use of the standard relations
\bea
&& F_D^\pm \ = \ \frac{1}{2} \left({\cal F}^\pm \ + \ i\, \widetilde{\cal F}^\pm \right) \ , \nonumber \\
&& F_A^\pm \ = \ \frac{1}{2} \left({\cal F}^\pm \ - \ i\, \widetilde{\cal F}^\pm \right) \ , \label{two_to_four}
\eea
in four--component notation the three quartic terms compatible with supersymmetry read
\bea
I_{++--} &=& \frac{1}{4} \left[ {\cal F}^+\cdot {\cal F}^+ \ {\cal F}^-\cdot {\cal F}^- \ + \ {\cal F}^+\cdot \widetilde{\cal F}^+\ {\cal F}^-\cdot \widetilde{\cal F}^- \right. \nonumber \\
&&\left. +\,  i \ {\cal F}^+\cdot \widetilde{\cal F}^+ \ {\cal F}^-\cdot {\cal F}^- \ - \ i\, {\cal F}^+ \cdot {\cal F}^+ \ {\cal F}^- \cdot \widetilde{\cal F}^- \right] \label{Ippmm}\\
I_{--++} &=& \frac{1}{4} \left[ {\cal F}^+\cdot {\cal F}^+ \ {\cal F}^-\cdot {\cal F}^- \ + \ {\cal F}^+\cdot \widetilde{\cal F}^+\ {\cal F}^-\cdot \widetilde{\cal F}^- \right. \nonumber \\
&&\left. -\,  i \ {\cal F}^+\cdot \widetilde{\cal F}^+ \ {\cal F}^-\cdot {\cal F}^- \ +\ i\, {\cal F}^+ \cdot {\cal F}^+ \ {\cal F}^- \cdot \widetilde{\cal F}^- \right] \label{Immpp} \\
I_{+-+-} &=& \frac{1}{4} \left[ {\cal F}^+\cdot {\cal F}^- \ {\cal F}^+\cdot {\cal F}^- \ + \ {\cal F}^+\cdot \widetilde{\cal F}^-\ {\cal F}^+\cdot \widetilde{\cal F}^- \right] \ . \label{Ipmpm}
\eea
Finally, if one demands the presence of an even number of ${\cal F}$ and $\widetilde{\cal F}$, as in the BI Lagrangians, only two combinations are left, $I_{+-+-}$ and the sum of the first two.

One can now verify that, while $I_{+-+-}$ reproduces the quartic term of the $U(1) \times U(1)$ model, the other combination does not reproduce the corresponding term of the $SU(2)$ model, due to first contribution present in both eqs.~\eqref{Ippmm} and \eqref{Immpp}. Similar considerations apply to the quartic terms of the $U(2)$ model, which also contains the peculiar last term in eq.~\eqref{lagfinu2}.
The indications for the $U(1) \times U(1)$ model are consistent with \cite{ABMZ}, since a BI of this type can be recovered, freezing the scalar, from the $N=1$ case of their generic $U(N,N)$ models, and supersymmetric versions were also given there. In superspace, the supersymmetric $U(1) \times U(1)$ model is indeed obtained replacing in \cite{CF} $W^\alpha \, W_\alpha$ with
\beq
W^{2\, +-}\equiv W^{+\alpha}\, W^-_\alpha \ ,
\eeq
so that the Lagrangian becomes of the form
\beq
{\cal L} \ = Re\int d^2 \theta  \ W^{2\, +-} \ + \ \int d^4 \theta \ W^{2\, +-}\ \overline{W}^{2\, +-} \ \Psi\left( D^2 \,W^{2\, +-} \,, \,\overline{D}^2\, \overline{W}^{2\, +-} \right) \ .
\eeq
where $\Psi$ is in principle an arbitrary function, to be adapted to the present case.
\section{Concluding remarks}\label{sec:4}

We have displayed three two--field extensions of the BI theory that realize the possible duality groups, namely $U(2)$, $SU(2)$ and $U(1) \times U(1)$. They were derived systematically from the IZ formalism and all rest on the same expression depending on a single auxiliary variable $a$,
\bea
{\cal L} \ =\ -\ \frac{2}{1\ -\ a}\,\left(\Re[\nu_t]\ +\ a\right)\ , \label{Lgena_concl}
\eea
while different definitions of $a$ give rise to the differences among the various cases:
\bea
&   a\ =\ {\rm Tr}\left( \overline{\nu}\, \nu \right) \ +\ 2\,\sqrt{{\rm Det}\left( \overline{\nu}\, \nu \right)}  \quad  & U(2)\ ;\\
&  a\ =\ {\rm Tr}\left( \overline{\nu}\, \nu \right)\ +\ 2\,\Re[{\rm Det}\,\nu ] \qquad   \quad  & SU(2)\ ; \\
& a\ =\ \left|{\rm Tr}\left({\nu} \right)\right|^2 \qquad   \quad   & U(1) \times U(1) \ .
\eea
Amusingly, the same type of expression entered, as we reviewed in Section \ref{sec:2}, a similar formulation of the standard BI theory that was first presented in \cite{BIK1}.

Passing to the ordinary field strengths ${\cal F}_{\mu\nu}$ of the four--component formalism, via the redefinitions \eqref{two_to_four} and their complex conjugates,
one obtains well--distinct forms for the three examples of Lagrangians. For the sake of brevity, let us now introduce complex combinations of the four--component field strengths, as in Section \ref{sec:susy},
\beq
{\cal F}^{+\,mn} \ = \ {\cal F}^{\,1\ mn} \ + \ i\, {\cal F}^{\,2\ mn} \ , \qquad {\cal F}^{-\,mn} \ = \ { \cal F}^{\,1\ mn} \ - \ i\, {\cal F}^{\,2\ mn} \ .
\eeq
The results that we have illustrated are then as follows (here we are not reinstating $f$):
\begin{itemize}
\item[1. ] \emph{Lagrangian with $U(1) \times U(1)$ duality:}
\end{itemize}
\beq\label{U1xU1}
{\cal L} \ =\ 1 \ -\ \sqrt{1\ +\ \frac{1}{2}\,\left({\cal F^+}\cdot {\cal F^-}\right) \ -\ \frac{1}{16}\,\left|{\cal F^+}\cdot\widetilde{{\cal F^-}}\right|^2} \ ;
\eeq
\begin{itemize}
\item[2. ] \emph{Lagrangian with $SU(2)$ duality:}
\end{itemize}
\beq\label{SU2}
{\cal L}\ =\ 1 \ -\ \sqrt{1\ +\ \frac{1}{2}\,\left({\cal F^+}\cdot {\cal F^-}\right) \ -\ \frac{1}{16}\,\left|{\cal F^+}\cdot\widetilde{{\cal F}^+}\right|^2} \ ;
\eeq
\begin{itemize}
\item[3. ] \emph{Lagrangian with $U(2)$ duality:}
\end{itemize}
\beq\label{U2}
{\cal L} \ =\ 1 \ -\ \sqrt{\left[1\ +\ \frac{1}{4}\,\left({\cal F^+}\cdot {\cal F^-}\right)\right]^2 \ -\ \frac{1}{32}\,C \ - \ \frac{1}{32}\sqrt{D}} \ ,
\eeq
where
\bea
C &=& \left|\left({\cal F^+}\right)^2\right|^2 \ +\ \left({\cal F^+}\cdot {\cal F^-}\right)^2\ +\ \left| {\cal F^+}\cdot\widetilde{{\cal F^-}}\right|^2\ +\ \left| {\cal F^+}\cdot\widetilde{ {\cal F^+}}\right|^2 \ , \\
D &=& \left[ \left( {\cal F}^+ \cdot {\cal F}^-\right)^2  \, - \, \left( {\cal F}^+ \cdot {\widetilde{\cal F}}^-\right)^2 \, + \, \left| {\cal F^+}\cdot\widetilde{ {\cal F^+}}\right|^2 \,  - \, \left|{\cal F^+}^2\right|^2 \right]^2  \\
&+& \left[ \left({\cal F^+}\right)^2 \, \left({\cal F}^- \cdot {\widetilde{\cal F}}^-\right) \, +\, \left({\cal F^-}\right)^2\, \left({\cal F}^+ \cdot {\widetilde{\cal F}}^+\right)
\, - \, 2\, \left({\cal F}^+ \cdot {\cal F}^-\right) \left({\cal F}^+ \cdot {\widetilde{\cal F}}^-\right) \right]^2 \ . \nonumber
\eea

As we have seen in Section \ref{sec:2}, in the single--field BI case ${\cal L}$ takes again the form in eq.~\eqref{Lgena_concl}, with $\nu_t$ replaced by $\nu$ and $a = \nu \,\overline{\nu}$. Moreover, in Section \ref{sec:2.2} we have displayed a one--parameter family of one--field models compatible with $U(1)$ duality, which also includes some unusual terms that are not analytic at the origin of field space. We built this simpler class of models since terms of a similar type also show up in our $U(2)$ example. Their emergence cannot be disentangled from the simplifying assumption of eq.~\eqref{conj}, which on the other hand was instrumental to arrive at a closed--form inversion. Clearly, we are not excluding that more conventional $U(2)$ solutions exist, but a closed--form inversion from IZ variables seems unlikely in more general cases. Our results should thus be contrasted with the earlier analysis in \cite{ABMZ}, which led to formal power--series presentations of models that apparently lack this peculiarity.

Energy positivity is clearly an important feature, which we are investigating further in these generalized BI constructions. While in the $U(1) \times U(1)$ and $SU(2)$ models positivity follows from the corresponding result for the BI theory, in the $U(2)$ example (or in its simpler one-field counterpart of eq.~\eqref{c}) it is less obvious. $U(2)$ duality ought to play a role in these considerations for the more complicated $U(2)$ model, but so far we have verified this key property only in a number of special field configurations, finding however no problems.

Finally, we have explained how the $U(1) \times U(1)$ model allows a straightforward ${\cal N}=1$ supersymmetric completion, which can be simply deduced from \cite{CF} replacing in the standard BI action $W^2$ with $W^{2\,+-}$, along the lines of what happens for its bosonic counterpart.

It would be interesting to explore point--like solutions in all these models with extended duality. The extension to $N$--field Lagrangians with $U(N)$ duality or subgroups thereof is another interesting problem. It would rest on generalizations of the invariants described here for the $N=2$ case, but no similar simplifications have emerged, so far, for $N>2$.
\vskip 24pt
{ The authors have had the privilege to contribute, with M.~Porrati, to the last paper of Raymond Stora \cite{FPSSY2}. Incompatible Academic commitments made it impossible to contribute together with Massimo, as we had originally planned, to this issue dedicated to the memory of Raymond.}
\vskip 24pt

\noindent{\large \bf Acknowledgements}\\ \noindent
We  are grateful to P.~Aschieri, I.~Antoniadis, S.~Bellucci, J.~Broedel, J.J.M.~Carrasco, B.L.~Cerchiai, E.~Dudas, R.~Kallosh, S.~Krivonos and M.~Porrati for discussions and/or collaboration on related issues. This work was supported in part by Scuola Normale Superiore, by INFN (I.S. GSS and Stefi) and by the ``Enrico Fermi Center''. A.~S. is grateful to the CPhT--\`Ecole Polytechnique, A.~Y. is grateful to Scuola Normale Superiore, while A.~S. and A.~Y. are both grateful to CERN, for the kind hospitality extended to them while this work was in progress.
%

\setcounter{equation}{0}

\end{document}